\documentclass[english,11pt,a4paper]{article}
\usepackage{amsmath,amssymb,amsthm} 

\usepackage{graphicx}
\usepackage[scanall]{psfrag}
\usepackage{comment}
\usepackage{tipa}

\newtheorem{hypothesis}{Hypothesis}

\title{Modal locking between vocal fold \\ and vocal tract oscillations: \\
experiments and statistical analysis}

\author{D.~Aalto$^1$,  J.~Malinen$^{2,3}$, and  M.~Vainio$^4$ \\ 
 {\small $^1$ Speech Communication and Disorders, University of Alberta, Canada} \\
 {\small $^2$ Dept. Mathematics and Systems Analysis, Aalto University, Finland } \\
 {\small $^3$ Dept. Signal Processing and Acoustics, Aalto University, Finland } \\
 { \small $^4$ Inst. Behavioural Sciences (SigMe group), University of Helsinki, Finland }
 \\ {\small \tt jarmo.malinen@aalto.fi }}

\date{\today}

\begin{document}

\maketitle

\begin{abstract}
The human vocal folds are known to interact with the vocal tract
acoustics during voiced speech production; namely a nonlinear
source-filter coupling has been observed both by using models and in
{\em in vivo} phonation. These phenomena are approached from two
directions in this article. We first present a computational dynamical
model of the speech apparatus that contains an explicit filter-source
feedback mechanism from the vocal tract acoustics back to the vocal
folds oscillations.  The model was used to simulate vocal pitch glides
where the trajectory was forced to cross the lowest vocal tract
resonance, i.e., the lowest formant $F_1$. A more detailed analysis
of the simulations is given in a companion article.

Similar vocal glide patterns produced by human participants are
studied in this article.  Both the simulations and the experimental
results reveal an effect when the glides cross the first formant (as
may happen in \textipa{[i]}).  Conversely, this effect is not observed
if there is no formant within the glide range (as is the case in
\textipa{[\textscripta]}). The experiments show smaller effect
compared to the simulations, pointing to an active compensation
mechanism.
\end{abstract}


\noindent {\bf PACS.} Primary 4370Bk. Secondary 4370Gr, 4370Aj, 4370Mn.


\newpage
\section{Introduction}

It is an underlying assumption in the linear source--filter theory of
vowel production that the sound source (i.e., the vocal fold
vibration) operates independently of the filter (i.e., the vocal
tract, henceforth VT) whose resonances modulate the resulting vowel
sound \cite{chika,fant:1960}.  This classical approach captures a wide
range of phenomena in speech production, at least in male
speakers. However, significant observations remain unexplained by the
feedback free source-filter model, such as some fine structure in the
phonation of a female singer near the lowest formant $F_1$.
Instability of the fundamental (glottal) frequency $f_0$ has been
detected acoustically \cite{trp:2008} and at tissue level
\cite{ZMHWH:OAIVVFTIPNSFCCS} when a singer performs an $f_0$-glide
over $F_1$ on a steady vowel.  As argued by Titze \cite{titze:2008},
the observed frequency jumps are due to a feedback coupling from the
vocal sound pressure back to the glottal pulse generation, denoted
there as the nonlinear source-filter coupling. Since $F_1$ usually
lies well above $f_0$ in male phonation, this phenomenon occurs
typically in female subjects when they are producing vowels with a low
$F_1$.  In laboratory experiments, Titze et al. found more
instabilities in male subjects, possibly because males have less
experience in suppressing unwanted instabilities \cite{trp:2008}.

The vocal source instabilities have been modeled by low-order
mass-spring systems. A two-mass vocal folds model, coupled with a
resonator tube, showed coupling related effects when the dimensions of
the tube were manipulated \cite{HFH:VISTI}. Tokuda et al.  simulated
vocal pitch glides using a four-mass model to analyze the interactions
between vocal register transitions and VT resonances
\cite{TZKH:BMRTRVTR}. We give a brief treatment of our model
simulations in this article; more detailed analysis using a more
comprehensive phonation model is given in the companion article
\cite{A-A-M-M-V:MLBVFVTOPI}.

For the current study, the $f_0$-$F_1$ cross-over phenomenon is
approached from two complementary directions: 1) by simulating
$f_0$-glides numerically on a steady vowel where $F_1$ is near $f_0$,
using a dynamical vowel model
that includes the filter-source feedback loop described above; and 2)
by carrying out a related vowel glide production experiment on female
test subjects.

\section{\label{VocalFoldsSec} The vocal folds model}

\subsection{Anatomy}

Before introducing the glottis model, some anatomic details are
reviewed, and physiological control mechanisms (actuated by muscles)
affecting the speech characteristics are explained. Such explanation
should be regarded as an idealization since the effect of a single
muscle contraction can to some extent be compensated by other muscles
involved. The glottis model is based on a further simplification of
this idealization.

\begin{figure} [!t]
  \centering
\hspace{1mm}
\includegraphics[width=4cm]{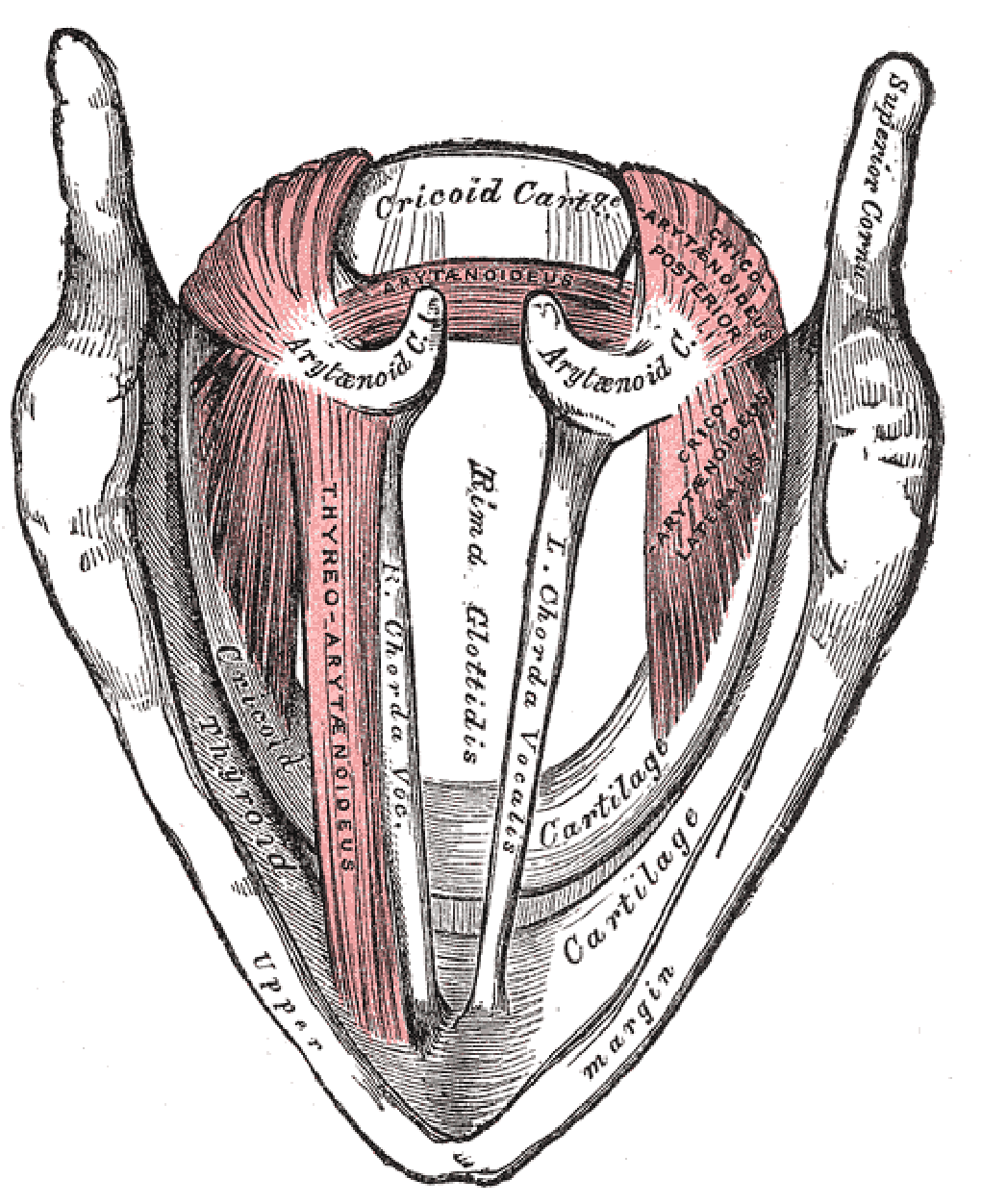} \hspace{5mm}
 \includegraphics[width=6cm]{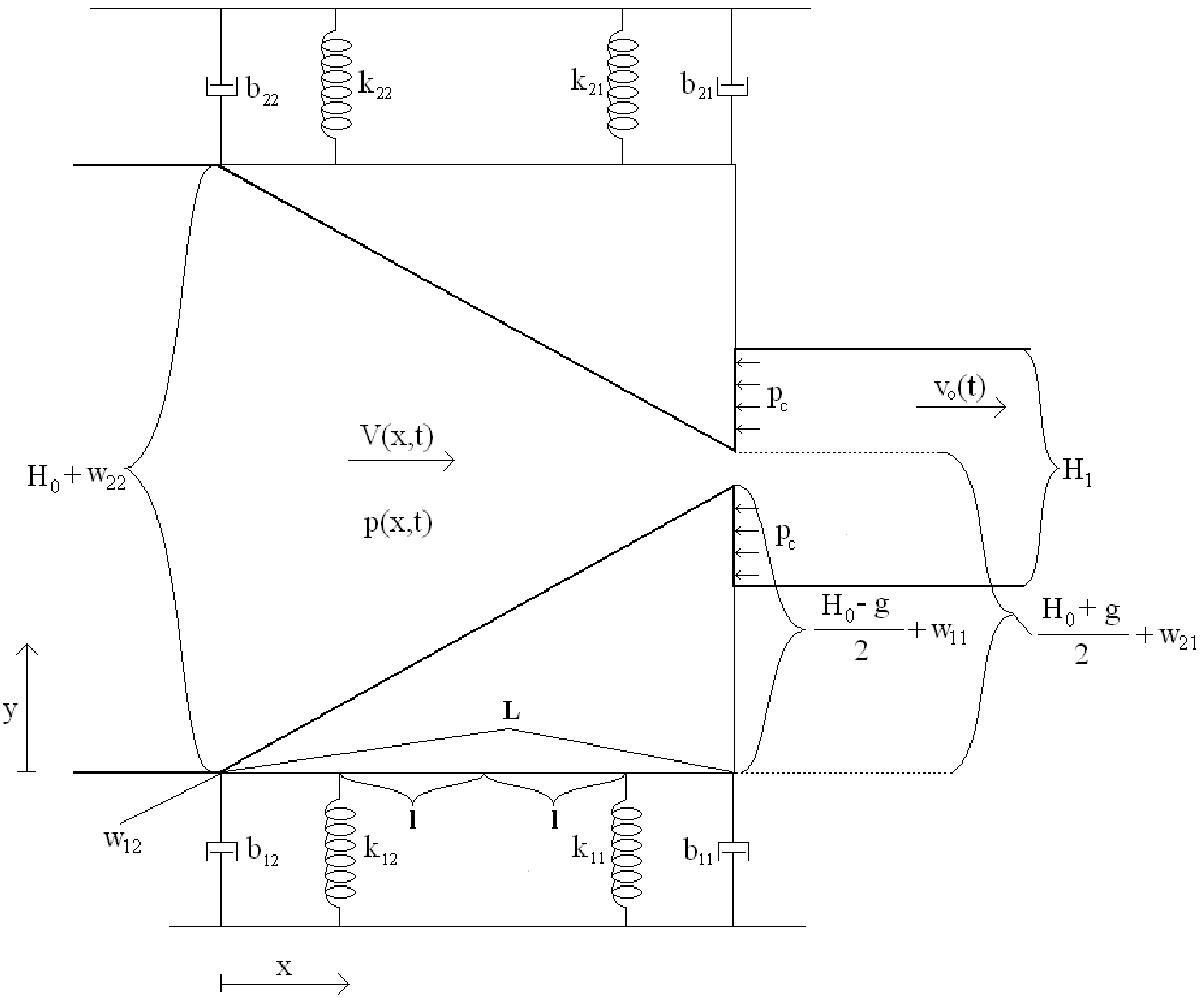} \\
 \hspace{-2cm} (a) \hspace{5.95cm} (b)
\caption{(a): Sketch of the anatomy of the glottis according to
Gray. (b): The geometry of the glottis model and the symbols
used. The trachea (i.e. the channel leading from the lungs to glottis)
is to the left in this sketch and the vocal tract is to the right.}
\label{fig:model}
\end{figure}

The vowel sound is produced by the self-sustained
cut-off effect of the air flow from lungs, caused by a quasi-periodic
closure of an aperture --- known as the rima glottidis --- by two
string-like vocal folds. This process is called phonation, and the
system comprising the vocal folds and the rima glottidis is known as
the glottis. A single period of the sound pressure signal, produced by
the glottal flow, is called the glottal pulse.

As shown in Fig.~\ref{fig:model}a, each vocal fold consists of a vocal
cord (also known as a vocal ligament) together with a medial part of
the thyroarytenoid muscle, and the vocalis muscle (not specified in
Fig.~\ref{fig:model}a). Both vocal folds are attached to the thyroid
cartilage from their anterior ends and to two corresponding arytenoid
cartilages (i.e., the left and the right) from their posterior ends;
see Fig.~\ref{fig:model}a.  In addition to these three cartilages,
there is the ring-formed cricoid cartilage whose location is inferior
to the thyroid cartilage. The vocal folds and the associated muscles
are supported by these cartilages as explained next.

Between each of the arytenoid cartilages and the cricoid cartilage,
there are two muscles attached. These are the posterior and the
lateral cricoarytenoid muscles, and they have opposite mechanical
actions. Indeed, during phonation the vocal folds are adducted by the
contraction of the lateral cricoarytenoid muscles, and conversely,
abducted by the posterior cricoarytenoid muscles during, e.g.,
breathing. This control action is realized by a rotational movement of
the arytenoid cartilages in a transversal plane. In addition, there is
a fifth (unpaired) muscle --- the arytenoid muscle --- whose
contraction brings the arytenoid cartilages closer to each other, thus
reducing the opening of the glottis independently of the lateral
cricoarytenoid muscles. These rather complicated control mechanisms
regulate, in particular, the type of phonation in the breathy-pressed
scale.

The fundamental frequency $f_0$ of a vowel sound is controlled by
another mechanism that is actuated by two cricothyroid muscles (not
visible in Fig.~\ref{fig:model}a). The contraction of these muscles
leads to a rotation of the thyroid cartilage with respect to the
cricoid cartilage.  Because of this rotation, the thyroid cartilage
inclines to the anterior direction, thus extending the vocal
folds. The elongation of the string-like vocal folds leads to increased stress
which raises the fundamental frequency of their longitudinal vibrations.

\subsection{Glottis model}

The anatomic configuration in Fig.~\ref{fig:model}a is modelled by a
low-order mass-spring system shown in Fig.~\ref{fig:model}b, based on
earlier work of Aalto \cite{Aalto:dt:2009,A-A-M-V:IVFVTO}.  Numerical efficiency and
theoretical tractability favor a low-order model, and the aim of the
model is at functionality rather than at anatomic detail.  The model
is able to reproduce accurately the measured male glottal pulse
obtained by inverse filtering \cite{lf:2009,A:GWAPSIAIF}.  The model
supports non-symmetric vocal fold vibrations, and both the fundamental
frequency $f_0$ as well as the phonation type can be chosen by
parameter values. Register shifts (e.g., from modal register to
falsetto) are important phenomena not in the scope of this
model. Accounting for register shifts would require either modelling
the vocal folds as aerodynamically loaded strings or by a high-order
mass-spring system that has a string-like ``elastic'' behavior.

The vocal fold model in Fig.~\ref{fig:model}b consists of two
wedge-shaped vibrating elements that have two degrees of freedom
each. The distributed mass of these elements can be reduced into three
mass points which are located so that $m_{j1}$ is at $x=L$, $m_{j2}$
at $x=0$, and $m_{j3}$ at $x=L/2$.  The elastic support of the vocal
ligaments is approximated by two springs at points $x=aL$ and $x=bL$.
The equations of motion for the vocal folds are given by
\begin{equation} \label{eq:liikeyhtalot}
\begin{cases}
  \  M_1\ddot{W}_1(t)+B_1\dot{W}_1(t)+K_1W_1(t)=-F(t), \\
  \ M_2\ddot{W}_2(t)+B_2\dot{W}_2(t)+K_2W_2(t)=F(t), \qquad t \in
  \mathbb{R}.
\end{cases}
\end{equation}
where $W_j=(w_{j1},w_{j2})^T$ are the displacements of the right and
left endpoints of the $j^{\textrm{th}}$ fold, $j = 1,2$.  The
respective mass, damping, and stiffness matrices $M_j$, $B_j$, and
$K_j$ in \eqref{eq:liikeyhtalot} are
\begin{equation} \label{eq:matrices}
\begin{array}{c}
M_j  =  \left[ \begin{array}{cc} m_{j1}+\frac{m_{j3}}{4} & \frac{m_{j3}}{4} \\
\frac{m_{j3}}{4} & m_{j2}+\frac{m_{j3}}{4} \end{array} \right],
\hspace{6mm} 
B_j = \left[ \begin{array}{cc} b_{j1} & 0 \\ 0 & b_{j2} \end{array} \right], \vspace{4mm}\\
\textrm{and } \quad K_j = P\left[ \begin{array}{cc} a^2k_{j1}+b^2k_{j2} & ab(k_{j1}+k_{j2}) \\ ab(k_{j1}+k_{j2}) & b^2k_{j1}+a^2k_{j2} \end{array} \right]. \end{array}
\end{equation}
The entries of these matrices are computed by means of Lagrangian
mechanics. The damping matrices $B_j$ are diagonal since the dampers
are located at the endpoints of the vocal folds. The springs are
located symmetrically around the midpoint $x=L/2$, so that
$a=(L/2+l)/L$ and $b=(L/2-l)/L$.  The control parameter $P > 0$ is
used for simulating variable frequency $f_0$-glides for the purpose of
this work.

During the glottal open phase (when $\Delta W_1(t) >0$), the load
terms of \eqref{eq:liikeyhtalot} are given by $F= (F_{A,1},F_{A,2})^T$
as given in Eq.~(\ref{eq:aeroforces}).  During the glottal closed
phase (when $\Delta W_1(t) <0$), there are no aerodynamic forces apart
from the acoustic counter pressure from the VT, denoted by $p_c$ and
properly introduced in the context of Eq.~\eqref{eq:webster}. Instead,
there is a nonlinear spring force for the elastic collision of the
vocal folds, given by the Hertz impact model \cite{H-S-S:NSSOHVFHMIF}:
\begin{equation} 
  F =\left[ \begin{array}{c} k_H|\Delta W_1|^{3/2}-\frac{H_0-H_1/2}{2L} \frac{H_1}{2}h \cdot p_c \\
      \frac{H_0-H_1/2}{2L} \frac{H_1}{2}h \cdot p_c
\end{array} \right].
\end{equation}

The model geometry is shown in Fig.~\ref{fig:model}b, and it
corresponds to the coronal section through the center of the vocal
folds.  As always in such biomechanical modelling \cite{TZKH:BMRTRVTR,
  H-S-S:NSSOHVFHMIF, H-S:glottis}, the lumped parameters of the
mass-spring system are in some correspondence to the masses, material
parameters, and geometric characteristics of the sound producing
tissues. Such parameters are, e.g., the mass and stiffness matrices
$M_i$ and $K_i$, $i=1,2$, that appear in the equations of motion for
the vocal folds \eqref{eq:liikeyhtalot}.

More precisely, matrices $M_i$ correspond
to the vibrating masses of the vocal folds, including the vocal
ligaments together with their covering mucous layers and (at least,
partly) the supporting vocalis muscles.  The elements of the matrices
$K_j$ are best understood as linear approximations of $k(s) = f/s$
where $f = f(s)$ is the contact force required for deflection $s$ at
the center of the string-like vocal ligament in Fig.~\ref{fig:model}a.
It should be emphasized that the exact numerical correspondence of
tissue parameters to lumped model parameters $M_j$ and $K_j$ is
intractable (and for practical purposes, even irrelevant), and their
values in computer simulations must be tuned using measurement data of
$f_0$ and the measured form of the glottal pulse \cite{lf:2009}.  Even
though the equations of motion are separate for both vocal folds, the
parameters (hence, the simulated vocal folds movements) are symmetric
in all of the simulations reported in this paper.

\section{\label{FullModelSec} Full model of vowel production}

Having treated the modelling of the vocal folds, it remains to review
the other components of the full model, i.e., the 1D incompressible
flow model and the VT acoustics model.

These two subsystems are coupled so that the flow depends on the
time-dependent glottal opening. Conversely, the flow produces
aerodynamic forces on the vocal folds, and it also acts as the
acoustic source to the resonating VT, modelled by Webster's equation.
The acoustics of the sub-glottal air cavities is not modelled at
all. The VT sound (counter) pressure gives rise to the filter-source
feedback from VT to the glottal oscillations. Without this feedback,
modal locking does not appear at all as can be verified by running
model simulations with $f_0 \approx F_1$ with $p_c = 0$ in
\eqref{eq:aeroforces}.

\subsection{Flow }

An incompressible 1D flow through the glottal opening with velocity
$v_o$ is described by
\begin{equation} \label{eq:Vout}
  \dot{v}_o(t)=\frac{1}{C_{iner}hH_1}\left(p_{sub}-\frac{C_g}{\Delta
      W_1(t)^3}v_o(t)\right)
\end{equation}
where the latter term inside the parentheses (representing the viscous
pressure loss) is motivated by the Hagen--Poiseuille law in a narrow
aperture.  The constant sub-glottal pressure (subtracted by the ambient
air pressure) is denoted by $p_{sub}$, and $h$ is the width of the
flow channel that is assumed rectangular. The parameter $C_{iner}$
regulates the flow inertia, and $C_g$ regulates the pressure loss in
the glottis.  Aalto et al. observed that $C_{iner}$ effectively
reveals the phonation type when other model parameters are estimated
based on recorded speech signals \cite{lf:2009}.

The viscous pressure loss in \eqref{eq:Vout} depends on the glottal
opening at the narrowest point, i.e., $\Delta W_1$.  At the other end
(i.e., towards the trachea) the opening is $\Delta W_2$.  These are
given by \eqref{eq:liikeyhtalot} through
\begin{equation}
\left[ 
 \begin{array}{c} \Delta W_1 \\
    \Delta W_2 \end{array}
\right]
=W_2-W_1+\left[
\begin{array}{c} g \\ H_0 \end{array} \right].
\end{equation}
The parameter $g$ is the glottal opening when there is no flow and the
vocal folds do not vibrate ($W_1=W_2=0$). In human anatomy, the
parameter $g$ is related to the position and orientation of the
arytenoid cartilages.

In the glottis, the flow velocity $V(x,t)$ is assumed to satisfy the
mass conservation law $H(x,t)V(x,t)=H_1v_o(t) $ for incompressible
flow where $H(x,t)$ is the height of the flow channel inside the
glottis. In the model geometry of Fig.~\ref{fig:model}b, we have
\begin{equation}
  H(x,t)=\Delta W_2(t)+\frac{x}{L}(\Delta W_1(t)-\Delta W_2(t)), \ \ \ x \in [0,L]. \nonumber
\end{equation}
Now the pressure $p(x,t)$ in the glottis is given by the two equations
above and the Bernoulli law $p(x,t)+\frac{1}{2} \rho V(x,t)^2=p_{sub}$
for static flow.

Since each vocal fold has two degrees of freedom, this pressure and
the VT counter pressure $p_c$ can be reduced to a force pair
$(F_{A,1},F_{A,2})^T$ where $F_{A,1}$ affects at the narrow (superior)
end of the glottis ($x=L$) and $F_{A,2}$ at the wide (resp. inferior)
end ($x=0$). This reduction is carried out by using the total force
and moment balance equations
\begin{equation} \label{eq:int1}
F_{A,1}+F_{A,2}=h \int_0^L (p(x,t)-p_{sub}) \, dx  \nonumber
\end{equation} 
and 
\begin{equation} \label{eq:int2}
L \cdot F_{A,1} =h \int_0^L x (p(x,t)-p_{sub}) \, dx-p_c \cdot h\frac{H_1}{2} \frac{H_0-H_1/2}{2}. \nonumber
\end{equation}
The moment is evaluated with respect to point $(x,y)=(0,0)$ for the
lower fold and $(x,y)=(0,H_0)$ for the upper fold in
Fig.~\ref{fig:model}b.  Evaluation of these integrals yields
\begin{equation} \label{eq:aeroforces}
    \begin{cases}
      \ F_{A,1} \ = \ \frac{1}{2} \rho v_o^2 h L \left(-
        \frac{H_1^2}{\Delta W_1(\Delta W_2-\Delta
          W_1)}+\frac{H_1^2}{(\Delta W_1-\Delta W_2)^2}
        \ln{\left(\frac{\Delta W_2}{\Delta W_1} \right)}\right)-
      \frac{H_1(H_0-H_1/2)}{4L} h p_c,    \vspace{2mm}  \\
     \ F_{A,2} \ = \ \frac{1}{2} \rho v_o^2 h L \left(
        \frac{H_1^2}{\Delta W_2(\Delta W_2-\Delta
          W_1)}-\frac{H_1^2}{(\Delta W_1-\Delta W_2)^2}
        \ln{\left(\frac{\Delta W_2}{\Delta W_1} \right) }\right)
      +\frac{H_1(H_0-H_1/2)}{4L} h p_c.
    \end{cases}
\end{equation}

\subsection{Vocal tract }

The vocal tract acoustics is modeled by Webster's lossless horn
resonator.  The governing equation for the velocity potential
$\Psi(s,t)$ is
\begin{equation} \label{eq:webster}
  \Psi_{tt}(s,t)-\frac{c^2}{A(s)}\frac{\partial}{\partial s} 
  \left( A(s) \frac{\partial \Psi (s,t)}{\partial s} \right)=0 
\end{equation}
where $c$ is the sound velocity.  The parameter $s \in [0,L_{VT}]$ is
the distance from the narrow (superior) end of the glottis measured
along the VT center line, and $L_{VT}$ is the length of the VT. The
area function $A(\cdot)$ is the cross-sectional area of the VT,
perpendicular to the VT center line.  The sound pressure is given in
terms of the velocity potential by $p=\rho \Psi_t$.

At the glottis end, the resonator is controlled by the flow velocity
$v_o$ from Eq.~\eqref{eq:Vout} through the boundary condition
$\Psi_s(0,t)=-v_o(t)$.  The resonator exerts a counter pressure
$p_c(t)=\rho \Psi_t(0,t)$ to the vocal folds equations
\eqref{eq:liikeyhtalot} through Eqs.~\eqref{eq:aeroforces}, thereby
forming a filter-source feedback loop. 
The boundary condition at lips is a frequency-independent acoustic
resistance of the form
$\Psi_t(L_{VT},t)+\theta c \Psi_s(L_{VT},t)=0$ 
where $\theta$ is the normalized acoustic resistance
\cite{Morse}. 

\subsection{Model parameters} 

The glottal flow equations \eqref{eq:Vout} contain two independent
parameters $p_{sub}/C_g$ and $C_{iner}$; the mass-spring system
contains three parameter matrices $M_j$, $B_j$, and $K_j$ for $j =
1,2$; and the resonator equations contain the area function $A(\cdot)$
in \eqref{eq:webster} and the acoustic termination parameter $\theta$
at the mouth.  

Out of these parameters, $p_{sub}/C_g$ and $\theta$ are determined
from physical considerations and $A(\cdot)$ from anatomical data
obtained by MRI.  The area function used in simulations corresponds to
\textipa{[\o:]} as in Hannukainen et al \cite{Hannukainen:VFW:2007}.
The parameter range for $C_{iner}$ has then been determined and
validated so as to produce a realistic time-domain glottal pulse form
\cite{lf:2009}; the values corresponding to pressed phonation are used
in this work. All these model parameter values introduced so far are
equally valid for both female and male phonation.

It remains to consider the parameter matrices in the vocal folds
equations \eqref{eq:liikeyhtalot} where the differences between female
and male phonation are significant.  Hor{\'a}{\v c}ek et al.  provide
parameter values in male phonation \cite{H-S:glottis,
  H-S-S:NSSOHVFHMIF}.  The parameter values ($K_1$, $K_2$) have been
estimated indirectly by requiring the correct fundamental frequency
$f_0$ \cite{Aalto:dt:2009}.  The focus of this paper is in female
phonation, and it is difficult to produce similar data for female
subjects using literature. Thus, the typical nominal ``male versions''
of parameters $M_j$ and $K_j$ for $j = 1,2$ (as given in Aalto
\cite{Aalto:dt:2009,lf:2009}) are scaled so as to obtain typical
``female version'' as explained above. This scaling is based on the
data given by Titze \cite{titze:1989}.

Dimensional analysis and scaling yield
\begin{equation} \nonumber M_{\textrm{female}}=\lambda^{\alpha}
  M_{\textrm{male}} \qquad \textrm{and} \qquad
  K_{\textrm{female}}=\lambda^{\alpha-2} K_{\textrm{male}}
\end{equation}
where the exponent $\alpha \in [2, 3]$ by physical grounds.  Based on
experimentation, the value used here is $\alpha=2.3$.  The scale
factor $\lambda \approx 0.6$ is suggested by Titze \cite{titze:1989}
but here the value $\lambda = 0.55$ produces more female like
phonation in simulations.  The resulting increase in the resonances of
\eqref{eq:liikeyhtalot} by factor $1/\lambda \approx 1.8$ reflects
correctly the higher pitch of the female voice.

The $f_0$-glide is simulated by additional scaling of the matrices
$K_j$ whereas the matrices $M_j$ are kept constant. This is based on
the assumption that the vibrating mass of vocal folds is not
significantly reduced when the speaker's pitch increases; a reasonable
assumption as far as register changes are excluded. The authors would
like to remark that the relative magnitudes of $M_j$ and $K_j$
essentially determine the resonance frequencies of model
\eqref{eq:liikeyhtalot}. However, attention must be paid to their
absolute magnitudes using, e.g., dimensional analysis since otherwise
the load terms $\pm F(t)$ in \eqref{eq:liikeyhtalot} (containing the
aerodynamic forces, contact force between the vocal folds during the
glottal closed phase, and the counter pressure from the VT) would
scale in an unrealistic manner.
 
The damping parameters $b_{ji}, \ i,j=1,2,$ in Eq.~\eqref{eq:matrices}
play an important but problematic role in glottis models. If there is
too much damping (while keeping all other model parameters fixed),
sustained oscillations do not occur. Conversely, too low damping will
cause instability in simulated vocal folds oscillations. The magnitude
of physically realistic damping in vibrating tissues is not available,
and the present model could possibly fail to give a quasi-stationary
glottis signal even if realistic experimental damping values were
available for use.  For simplicity, we set $b_{ji} = \beta > 0$ for
$i,j=1,2$, and the value of glottis loss $\beta$ is adjusted
separately for each parameter value set in order to obtain stable but
sustained oscillation.  In particular, parameter $\beta$ is adjusted
every $100 \ ms$ during the $f_0$-glide simulations presented
below. The damping remains always so small that its lowering effect on
the resonances of the mass-spring system \eqref{eq:liikeyhtalot} is
negligible.

Let us conclude by discussing the parameter magnitudes in
Eq.~\eqref{eq:liikeyhtalot}. The total vibrating mass is
$m_{j1}+m_{j2}+m_{j3}=0.48 \ g$ for male and $0.12 \ g$ for female
phonation. The total spring coefficients are $k_{j1}+k_{j2}=193 \ N/m$
for male and $161.3 \ N/m$ for female phonation using the nominal
values, i.e., when $P=1$ in Eq.~\eqref{eq:matrices}.  The nominal
values yield $f_0=110$ Hz for male and $187$ Hz for female
phonation. If the characteristic thickness of the vocal folds is
assumed to be about $5 \ mm$, these parameters yield a magnitude
estimate for the elastic modulus of the vocal folds by $E \approx
\frac{k_{j1}+k_{j2}}{Lh} \cdot 5 \cdot 10^{-3} \ m \approx 6.6
\ kPa$. This is in good comparison with Fig.~7 in Chhetri et
al.\cite{Chhetri:modulus} where estimates are given for the elastic
modulus of \emph{ex vivo} male vocal folds between $2.0 \ kPa$ and
$7.5 \ kPa$ for different parts of the vocal folds.

\subsection{Numerical realization}

The model equations are solved numerically using MATLAB software and
custom-made code. The vocal fold equations of motion
\eqref{eq:liikeyhtalot} are solved by the fourth order
Runge--Kutta time discretization scheme. The discontinuity of the load
$F(t)$ at $\Delta W_1(t)=0$ is dealt with by an interpolation
procedure detailed in Aalto (Section 2.4)\cite{Aalto:dt:2009}. The
flow equation \eqref{eq:Vout} is solved by the backward Euler method.
The VT is discretized by the FEM using piecewise linear elements
($N=100$) and the physical energy norm of Webster's equation.
Crank--Nicolson time discretization is used, and the time step is
always $20 \ { \mu s}$.

\section{\label{SimSec} Simulation results}

Frequency glides of vowel \textipa{[\o:]} are simulated near the
lowest formant $F_1$ or its subharmonic $F_1/2$. In these simulations,
$F_1$ (determined from spectrograms of simulated vowel signals)
coincides with the lowest resonance of the VT (solved independently
from the eigenvalue problem associated to \eqref{eq:webster} and the
boundary conditions). To produce the glides, parameter $P$ in
Eq.~\eqref{eq:matrices} is increased (or decreased) as a quadratic
function of time. Then the oscillation frequency of the vocal fold
model \eqref{eq:liikeyhtalot} in the absence of the counter pressure
$p_c$ --- denoted by $\tilde f_0$ --- increases (respectively,
decreases) as a linear function of time. This can be observed in
simulation results if the filter-source feedback mechanism is
disconnected by setting the counter pressure $p_c = 0$ in
Eq.~\eqref{eq:aeroforces}.  When using $p_c(t) =\rho \Psi_t(0,t)$ with
Eq.~\eqref{eq:webster}, the observed fundamental frequency $f_0$ of
the whole system does not behave linearly in time (in contrast to
$\tilde f_0$) but exhibits jumps near $F_1$ (and $F_1/2$) as shown in
Fig.~\ref{fig:sweeps1}b.  This behavior, the modal locking, is due to
the filter-source feedback \cite{titze:2008,ZMHWH:OAIVVFTIPNSFCCS}.

Changing the area function $A(\cdot)$ in Eq.~\eqref{eq:webster} to
correspond some other vowel than \textipa{[\o:]} does not change the
results of model simulations apart from formant, and hence, modal
locking frequencies.

\subsection{$f_0$--$F_1$ crossover}

Frequency $f_0$-glides of length 2 s are simulated by varying the
parameter $P$ of the glottis model so that 350 Hz $ < \tilde f_0 < $
810 Hz linearly in time.  The increasing phase is shown in the
spectrogram Fig.~\ref{fig:sweeps1}a with an auxiliary line
showing the glide of $\tilde f_0$ and another line showing $F_1 = 647$
Hz. It is observed that $f_0$ coincides first with $\tilde f_0$, but
then it suddenly jumps upwards to $F_1$ when it reaches about 470 Hz.
The wave form of the glottal pulse near the transition is a
superposition of two signals with frequencies $\tilde f_0$ and $F_1$.
When $\tilde f_0$ exceeds $F_1$, then $f_0$ and $\tilde f_0$ coincide
again.  The qualitative behavior is sketched in the rising part of
Fig.~\ref{fig:sweeps1}b.

In a downwards glide the behavior is almost symmetric. First $f_0 =
\tilde f_0$ descends down to $F_1$ where $f_0$ locks for a while.  The
latent glottal model frequency $\tilde f_0$, of course, goes down
linearly without change.  After a time lag, $f_0$ is released from
$F_1$ and drops suddenly to $\tilde f_0$. This behavior is shown in
the descending part of Fig.~\ref{fig:sweeps1}b.

As can be seen in Fig.~\ref{fig:sweeps1}b, there are two periods (of
length $T_u, T_d > 0$) during which $f_0 = F_1$. Let us denote the
corresponding frequency jumps by $\phi_u, \phi_d > 0$.  By
Fig.~\ref{fig:sweeps1}b we see $\phi_u/ T_u = \phi_d/ T_d$ is
determined by the ascend and descend rate of the simulated, symmetric
$\tilde f_0$-glide.  We observe $\phi_d > \phi_u$ consistently in all
simulations, and if the $\tilde f_0$-glide is very slow, we observe
that $\phi_d \approx \phi_u$ and Fig.~\ref{fig:sweeps1}b becomes
symmetric. Simulating extremely slow glides (during which the
parameter $P$ can be regarded as a constant), it is observed that the
jumps $\phi_d$ and $\phi_u$ have a common lower bound $\phi \approx$
174~Hz. This number can be regarded as the ``true'' magnitude of the
simulated frequency jump, and it is in reasonable correspondence with
the results reported in \cite{trp:2008,titze:2008}.

It was noted above that $\phi_d > \phi_u$. One way to understand this
asymmetry is in terms of the energy dissipation from VT resonance
modes. The only energy losses in Eq.~\eqref{eq:webster} are due to the
dissipative boundary condition at lips. During a downwards glide,
there is a locking of $f_0$ at $F_1$, and then a lot of energy is
contained in the corresponding eigenmode of the joint system,
comprising the VT air column and the vocal fold masses. For the vocal
folds oscillations to get ``unlocked'' from this eigenmode, most of
this energy must first be dissipated either by mouth radiation or by
dispersion to other eigenmodes as a consequence of nonlinearity and
lack of time invariance of the model.  The dissipation rate may be
slow compared to the speed of decrease of the $\tilde
f_0$-glide. Thus, the latent frequency $\tilde f_0$ may have fallen
far below $F_1$ before unlocking of $f_0$ may take place.  It is
therefore expected that the time delay $T_d$ is sensitive to relative
magnitudes of losses and amounts of energy stored in vocal folds and
VT, and this is supported by the simulations. Indeed, $T_d$ becomes
very large if the vocal folds oscillation amplitude is small during
the glide.

\begin{figure} [!t]
\vspace{-3mm}
\begin{center} 
 \includegraphics[width=6cm]{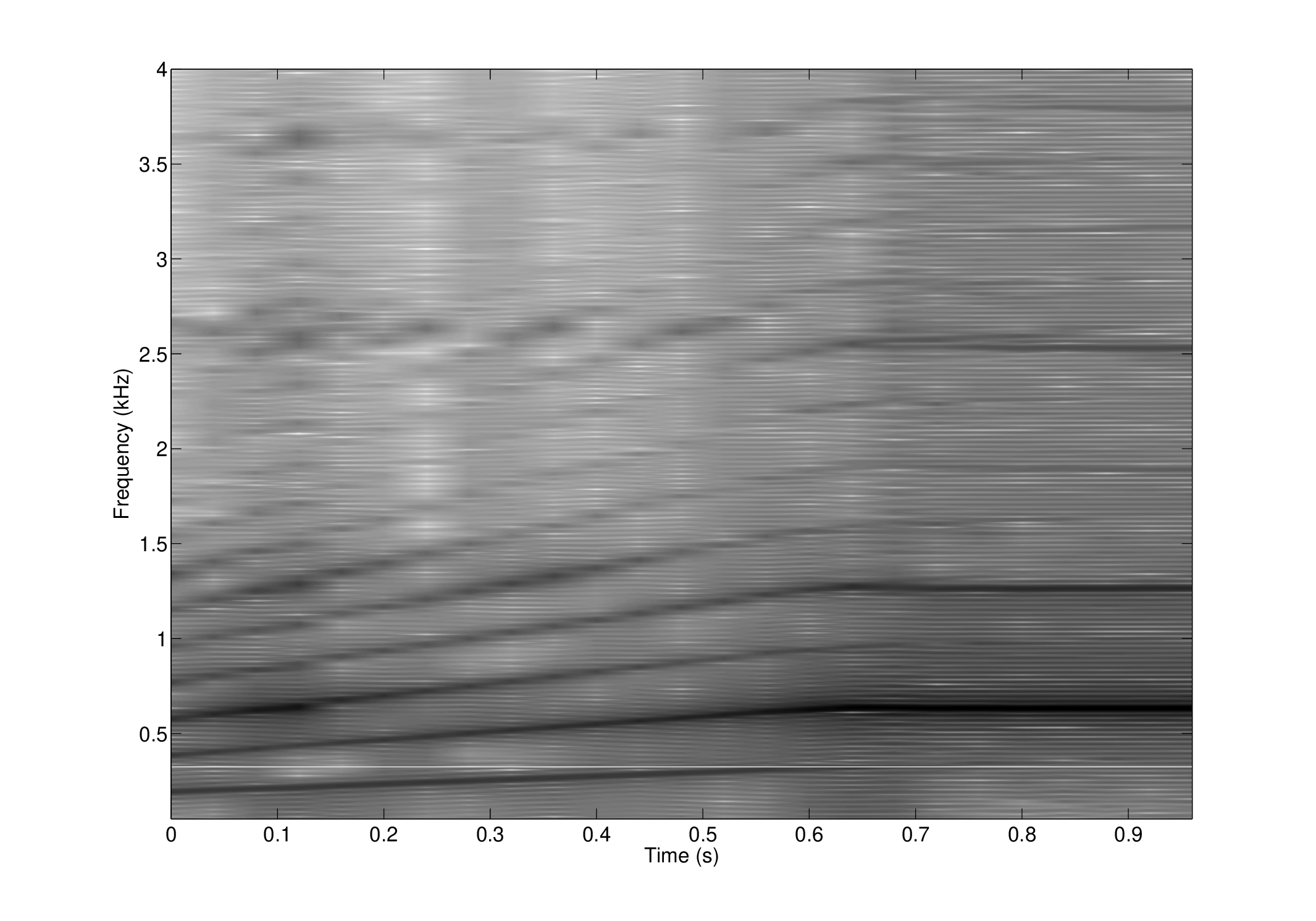} \hspace{-9mm}
 \includegraphics[width=6cm]{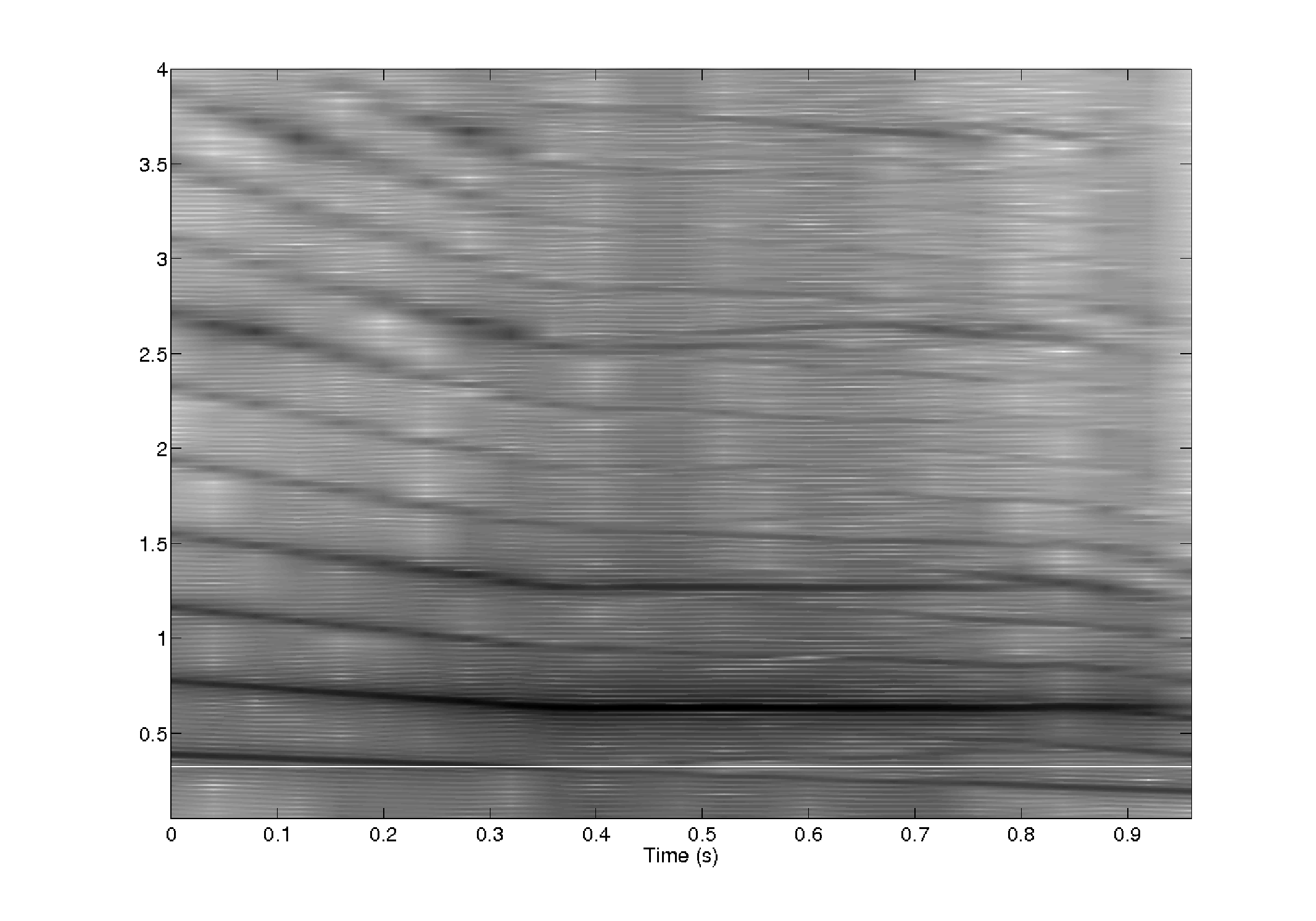}
\vspace{-4mm}
\end{center}
\caption{(a): Upwards $f_0$-glide over $F_1/2$. (b): Downwards $f_0$-glide over
  $F_1/2$.}
\label{fig:sweeps2}
\end{figure}

\subsection{Glides near subharmonics of $F_1$}

Similar $f_0$-glides are produced as explained above but now 187~Hz $
< \tilde f_0 < $ 390~Hz linearly during 2 s time interval.  The
subharmonic $F_1/2 =324$ Hz lies in this interval, and two kinds of
behavior are observed. First, $f_0$ increases from below in
Fig.~\ref{fig:sweeps2}a until $F_1/2$ is reached; there is no locking
at $F_1/2$ but $f_0$ jumps one octave up and locks at $F_1$. 
Second, the phenomenon does not appear at all if the vocal fold oscillation amplitude is large enough.

In Fig.~\ref{fig:sweeps2}b there is a downward glide near $F_1/2$.
First and third harmonics of $f_0$ have a static part matching $F_1$
and $2F_1$ respectively.  It may be possible that $f_0$ is ``partly
locked'' to $F_1/2$ but there is a clear component that develops with
$\tilde f_0$.  Such a complicated spectral behavior emphasizes the
surprising dynamical richness of the vowel production model presented
in this paper.
 
\section{Glide production experiment}

The simulation results and earlier experimental evidence
\cite{trp:2008} suggest that consequences of filter-source feedback
should be detectable at crossings of frequencies $f_0$ and
$F_1$. Unfortunately, determining the precise crossing time of $f_0$
and $F_1$ in vowel glides is difficult when based on the acoustic
signal alone. Indeed, the harmonics $2f_0, 3f_0, \ldots$ do not
coincide with the bandwidth of $F_1$ if $f_0 \approx F_1$. If $f_0$
follows successfully a slowly varying reference glide (as should happen
in these experiments), it is then far from a persistently exciting
signal that would be required for precise detection of $F_1$.  In a
similar glide production experiment, Titze et al. \cite{trp:2008}
asked the test subjects to produce a spectrally rich vocal fry
adjacent to every glide. A good estimate for $F_1$ can be obtained
this way, but the formant may creep during the glide. The subject is,
in fact, expected to use various techniques to avoid modal locking in
order to produce audibly clean vowel glides, and $F_1$ position is
expected to be affected near $f_0$.  We conclude that obtaining the
crossing time of $f_0$ and $F_1$ from acoustic signals remains a difficult
estimation problem.

\begin{figure} [!t]
\vspace{-3mm}
\begin{center} 
 \includegraphics[width=6.5cm]{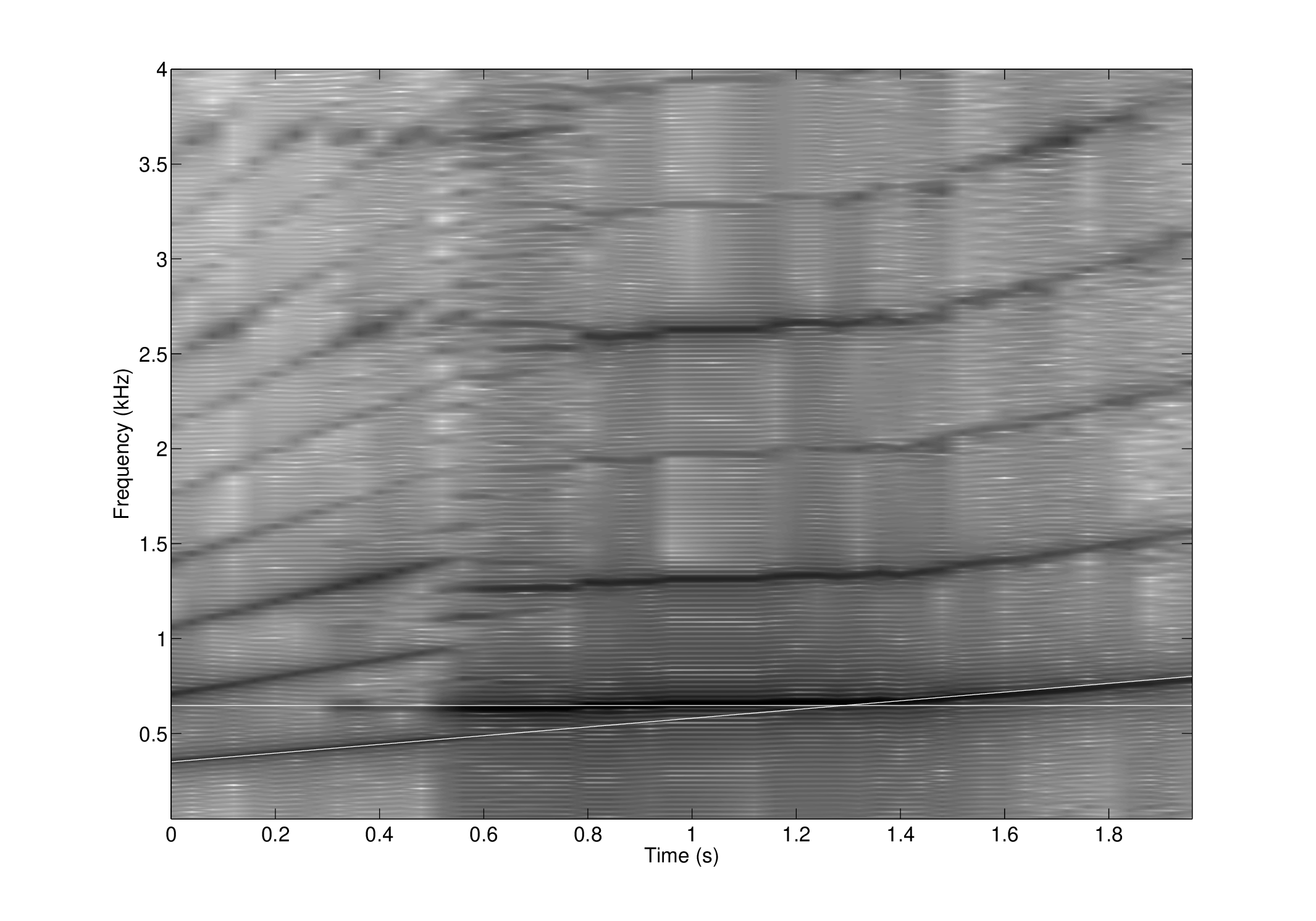} \hspace{-9mm} 
 \includegraphics[width=6cm]{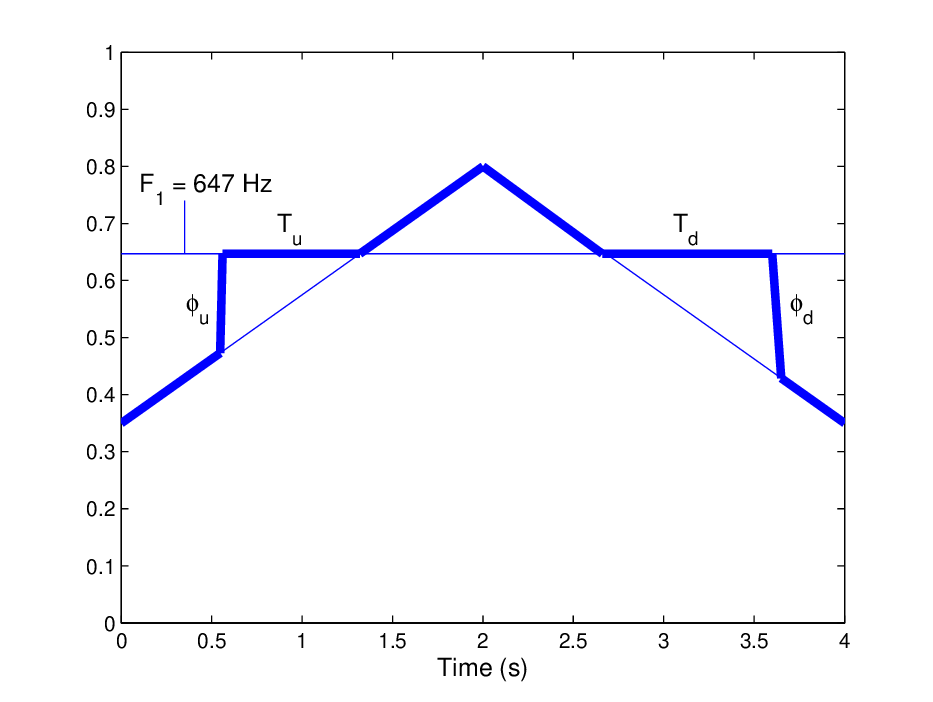} 
\vspace{-4mm}
\end{center}
\caption{(a): $f_0$-glide 350 Hz -- 810 Hz. (b): A sketch of the modal
  locking in an $f_0$-glide over $F_1$ first upwards and then
  downwards. The thick (thin) line shows $f_0$ (resp. $\tilde f_0$)
  during the glide.}
\label{fig:sweeps1}
\end{figure}

Rather than replicating the experimental arrangements of Titze et
al. \cite{trp:2008}, a complementary approach is taken here without
$F_1$ as a controlled variable.  This is achieved by eliciting vocal
glides for two different vowels \textipa{[\textscripta]} and
\textipa{[i]}, where \textipa{[i]} has $F_1$ within the $f_0$ glide
range, and $F_1$ of \textipa{[\textscripta]} is as far as possible
from the same range.  The model simulations predict that the
$f_0$-trajectories corresponding to \textipa{[\textscripta]} and
\textipa{[i]} should have distinctly different characteristics that
should be detectable in natural glide productions.

\subsection{\label{QuantSec} Quantification of the jumping pattern }

During a modal locking episode, $f_0$ is expected to jump as in
Fig.~\ref{fig:sweeps1}. To quantify the jump effect for statistical
analysis, the Duration Ratio ($\mathop{DR}$) of the glide $f_0 =
f_0(t)$ is defined.  For frequencies $f_A < f_a < f_b < f_B$,
$\mathop{DR}$ is defined by
\begin{equation} \label{gpi}
\mathop{DR}(f_0) = \frac{t_b-t_a}{t_B-t_A}
\end{equation}
where $(t_A,t_B)$ is the longest open interval where $f_A \leq f_0(t)
\leq f_B$ for all $t\in (t_A,t_B)$, and $(t_a,t_b)$ is the longest
open interval where $f_a \leq f_0(t) \leq f_b$ for all $t\in
(t_a,t_b)$ such that $[t_a,t_b] \subset (t_A,t_B)$, i.e., $t_A<t_a\leq
t_b<t_B$. This implies $0 \leq \mathop{DR}(f_0) \leq 1$ for any glide
$f_0$. By definition, 
$\mathop{DR}$ does not depend on the direction or the speed of the glide
in the sense that $\mathop{DR}(f_0) = \mathop{DR}(\tilde f_0) =
\mathop{DR}(f_0^a)$ where $\tilde f_0(t) = f_0(-t)$ and $f_0^a(t) =
f_0(at)$ for $a > 0$.
 
In the current experiments, all octave glides are between $200 $ Hz
and $400 $ Hz, and the parameter values for Eq.~\eqref{gpi} are always
$f_A = \gamma \cdot 200$ Hz $ \approx 230$ Hz, $f_a = \gamma f_A
\approx 264$ Hz, $f_b = \gamma f_a \approx 303$ Hz, and $f_B = \gamma
f_b \approx 348$ Hz where $\gamma=2^{1/5} \approx 1.148698$.  Hence,
the octave $[200 \, \mathrm{Hz}, 400 \, \mathrm{Hz}]$ is divided into
five intervals equal in logarithmic scale.  We say that the glide
$f_0$ is full if its range contains all of the interval $[f_A, f_B]$.
If $f_0(t) = a2^{kt}$ for some $a>0$, $k \neq 0$, and $t \in [0,T]$ is
a full glide, then we call it logarithmically linear and we have $
\mathop{DR}(f_0) = 1/3$ which can be regarded as the nominal value of
$\mathop{DR}$ on full glides in the absence of any perturbations.

For an ideal logarithmically linear full glide $f_0$ with a single
modal locking ``jump'' as in Fig.~\ref{fig:sweeps1}, we have
$\mathop{DR}(f_0) < 1/3$ if the jump intersects $f_b$ but not $f_a$ or
$f_B$.  Similarly, $\mathop{DR}(f_0) > 1/3$ if the jump intersects
$f_a$ but not $f_b$ or $f_A$. The jump size and the jump position
$\tilde F$ (satisfying $\tilde F = F_1$ in glide simulations) change
the $\mathop{DR}$ in a practically convenient way: to see this, a
numerical experiment was performed.  Full, logarithmically linear jump
patterns were constructed using jump size of $100$ Hz and one million
tokens of $\tilde F$ drawn from a normal distribution with $\mu = 300$
Hz and $\sigma = 50$ Hz. This produced $\mathbb{E}[\mathop{DR}] =
0.308$ and $\mathop{SD}[\mathop{DR}] = 0.325$.  Keeping $\sigma$
constant but varying $\mu$ gave the following values:
$\mathbb{E}[\mathop{DR}] = 0.354$ and $\mathop{SD}[\mathop{DR}] =
0.270$ at $\mu = 250$ Hz; $\mathbb{E}[\mathop{DR}] = 0.240$ and
$\mathop{SD}[\mathop{DR}] = 0.270$ at $\mu = 350$ Hz; and finally
$\mathbb{E}[\mathop{DR}] = 0.333$ and $\mathop{SD}[\mathop{DR}] =
0.002$ at $\mu = 600$ Hz as expected.

The experiment is designed so as to create situations where glides of
\textipa{[i]} are more likely to have lower $\mathop{DR}$ compared to
glides of \textipa{[\textscripta]}, due to modal locking induced
$f_0$-jumps over $f_b$ for \textipa{[i]} but not at all for
\textipa{[\textscripta]}.  More precisely, the position of $F_1$ of
\textipa{[i]} is expected to be roughly at $300$ Hz (which, of course,
explains the choice of $f_b$ above), and the position of $F_1$ of
\textipa{[\textscripta]} is likely to be above $600$ Hz.  Thus,
simulated model predictions of jump patterns can be formulated as the
statistically testable hypotheses:
\begin{hypothesis} \label{Hypo1}
  The population mean of the Duration Ratio on full
  \textipa{[i]}-glides $f_0^{\text{\textipa{[i]}}}$ and on full
  \textipa{[\textscripta]}-glides
  $f_0^{\text{\textipa{[\textscripta]}}}$ satisfies $\mathbb{E}\left [
    \mathop{DR}(f_0^{\text{\textipa{[i]}}}) \right ] < \mathbb{E}\left
          [\mathop{DR}(f_0^{\text{\textipa{[\textscripta]}}}) \right
          ]$.
\end{hypothesis}
\begin{hypothesis} \label{Hypo2}
  The population mean of the Duration Ratio on full
  \textipa{[i]}-glides $f_0^{\text{\textipa{[i]}}}$
satisfies
  $\mathbb{E}\left [ \mathop{DR}(f_0^{\text{\textipa{[i]}}}) \right ] < 1/3$, 
and the population mean of the Duration Ratio on full 
\textipa{[\textscripta]}-glides $f_0^{\text{\textipa{[\textscripta]}}}$
satisfies
  $\mathbb{E}\left [\mathop{DR}(f_0^{\text{\textipa{[\textscripta]}}}) \right ] = 1/3$.
\end{hypothesis}

\subsection{Subjects and the recording arrangement}

Eleven native Finnish speaking female students at the University of
Helsinki were recruited for the experiment.  Females were chosen as
subjects because (non-singing) males have typically little familiarity
of using their vocal range around their naturally occurring $F_1$.
Hence, using males would probably lead to an overly high rejection
rate of glide samples and excursions to falsetto register in otherwise
successful glides.

None of the participants reported any hearing or voice problems, and
they had not received professional training in singing although some
of them had a musical hobby (like violin playing).  The participants
were informed of the general purpose of the research and presented
with the gliding task. There was a short familiarization period before
the actual experiment during which the participants could practice the
gliding with the vowel \textipa{[\textscripta]}.  Some participants
were helped to find the right pitch range.  The data was recorded in a
sound-proof studio with a high-quality microphone (AKG C4000B), and
digitized with Digidesign (DIGI 002) and ProTools v.9.

\subsection{Glide imitation stimuli}

The subjects were instructed to produce frequency glides by following
a spectrally neutral, synthesized reference glides from 200 Hz to 400
Hz. Every reference glide started and ended with a constant frequency
part of $0.25$ s to create a sensation of the right initial (resp.,
final) pitch of the glide.  The instruction was given as a
pre-recorded 0.5 s sample of either of the vowels, followed by three
beeps (countdown beep: 0.25 s signal followed by 0.75 s silence).  The
beeps had the pitch of the onset value of the desired glide (either
200 Hz or 400 Hz).  The glides and beeps were frequency modulated
triangle waves; the triangle wave function $s(t)$ is a $2\pi$-periodic
function with constant slope (except for multiplicities of $2\pi$).
The upward glides are given by $s(2\pi \omega (t))$ with $ \omega(t) =
\frac{T\cdot 200 \mathrm{Hz}}{\log 2}(2^{t/T}-1)$ for $ t\in [0,T]$
where $T$ is the duration of the glide (either $1.5$ s or $3.0$ s).
The instantaneous frequency of such a glide is given by
$\frac{\partial\omega(t) }{\partial t} = 2^{t/T} \cdot 200
\mathrm{Hz}$ which was assumed to be equal with the perceived pitch.

\subsection{Experimental setup}

The subjects were asked to imitate the pitch of reference glides by
producing them with a prolonged, spoken \textipa{[\textscripta]} or
\textipa{[i]}.  The subjects heard the reference glide from earphones
at a volume that was adjusted for each subject to be as loud as
possible without being unpleasant.  The purpose of this arrangement
was to hinder the subjects from hearing their own production easily 
and to elicit good phonation.

The data was gathered following a factorial $2^3$ design with
direction (either fall or rise), duration (either fast 1.5 s or slow 3
s), and vowel (either \textipa{[\textscripta]} or \textipa{[i]}) as
factors.  Each subject had her own pseudo-randomized list of
stimuli. The lists were constructed so that the different stimulus
types would be randomly and evenly distributed in the trial: the
stimuli were divided in blocks of 24 where each stimulus condition
occurred three times, three identical stimuli were never in a row (not
even over the blocks), four identical glide directions, vowels, or
glide speeds were never in a row.

\subsection{Data analysis}

Three subjects were disqualified since they did not learn the gliding
pattern well enough. Most of their productions had at least one of the
following characteristics: the glide was one octave away from the
guiding glide signal, the glide pattern was missing (static or
undulating pitch), or the range of the glide was limited to less than
a quarter.  The remaining eight subjects had no difficulties in the
gliding task.

A total of 864 glides were processed further.  The vocal pitch was
extracted by cross-correlation using Praat (Praat 5.2.21; default
parameters with admissible range from 200 Hz to 400 Hz).  The pitch
trajectories of the glides were automatically chopped from the audio
signal, based on the timing of the guiding signals and further
analyzed in MATLAB.  A custom-made algorithm was used to calculate
$\mathop{DR}(\cdot)$ from the data.

The glides were analyzed for their ``fullness'' to exclude the related
bias in Duration Ratio values. The range from $f_A$ to $f_B$ was
divided in ten equally long frequency bands in logarithmic scale. If
some of the bands did not contain a value of $f_0$, the glide was not
considered full.  This led to exclusion of 8 glides (1\%).  A mixed
linear model was fitted to the remaining data using statistical
software package R (version 2.14.0) and the model parameter selection
was based on log-likelihoods \cite{pinheirobates}. Both of the
Hypotheses~\ref{Hypo1} and \ref{Hypo2} were independently tested by
t-tests.

\begin{figure} [!t]
  \centering
\hspace{1mm}
\includegraphics[width=8cm]{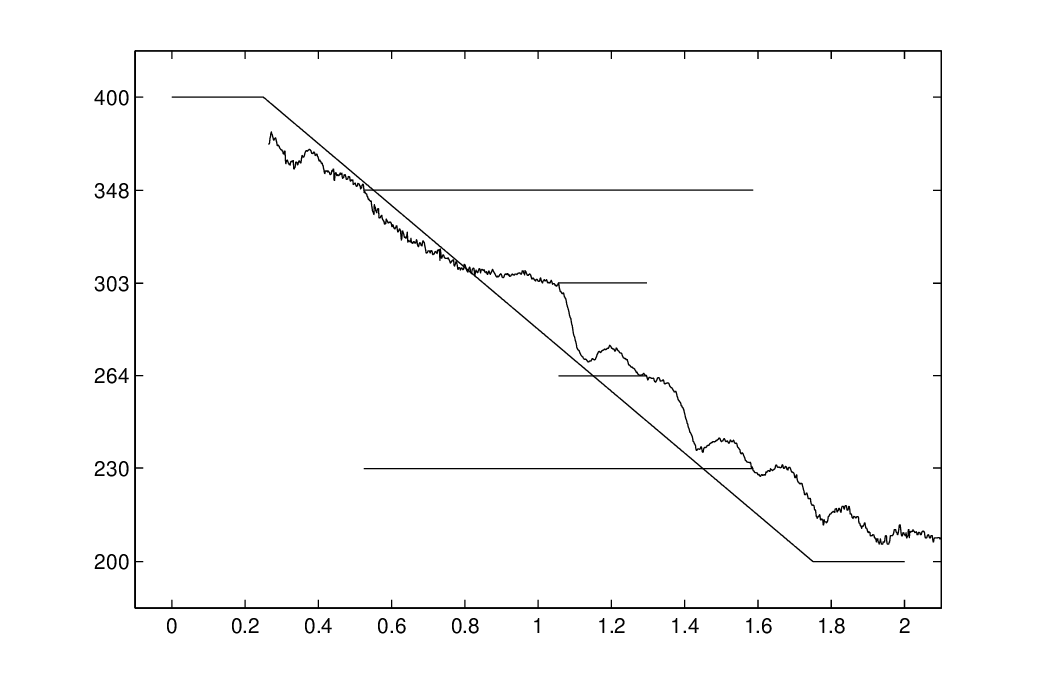}
\caption{A typical production of a vowel glide \textipa{[i]}. 
The linear line depicts the guiding signal
and the zig-zagging line shows the fundamental frequency of the glide. The intervals
corresponding to the mid-frequency and total durations are shown as horizontal lines.
}
\label{fig:i_glide}
\end{figure}

A typical production is shown in Fig.~\ref{fig:i_glide} where the
$f_0$ trajectory of a fast falling \textipa{[i]} is given.  The
reference glide is the diagonal line with flat ends at $t = 0, 2$. The
four horizontal lines indicate the critical frequencies $f_A, f_a,
f_b$ and $f_B$ as introduced above. The corresponding time intervals
for Eq.~\eqref{gpi} are given by $(t_A,t_B) = (0.52,1.59)$ and
$[t_a,t_b] = (0.77,1.03)$, yielding $\mathop{DR}(f_0) = 0.243 <
1/3$.  

\subsection{Results}

  The estimated means and standard deviations for $\mathop{DR}$ at each eight
  factor combination groups are shown in the Table~\ref{tableDR}. The
  estimated mean $\mathop{DR}$'s of \textipa{[\textscripta]} are consistently
  larger than those of \textipa{[i]} (see Hypothesis~\ref{Hypo1}), and
  also the nominal $\mathop{DR}$ value $1/3 = 0.33\ldots$ is above the
  estimated mean of $\mathop{DR}$ for all four \textipa{[i]} conditions.

\begin{table}[t,h]
\caption{\it The means and standard deviations for the Duration Ratio
  ($\mathop{DR}$) over each factor combination.}
\vspace{2mm}
\centerline{
\begin{tabular}{|c|c|c|c|}
\hline
Condition & Mean & SD & N\\
\hline
Fast fall \textipa{[\textscripta]} & 0.341 & 0.099 &106\\
Fast fall \textipa{[i]} & 0.324&0.108&105\\
Fast rise \textipa{[\textscripta]} & 0.331 & 0.084&108\\
Fast rise \textipa{[i]} & 0.312 & 0.073&108\\
Slow fall \textipa{[\textscripta]} & 0.337 & 0.106&107\\
Slow fall \textipa{[i]} & 0.326 & 0.104&106\\
Slow rise \textipa{[\textscripta]} & 0.340 & 0.102&108\\
Slow rise \textipa{[i]} & 0.320 & 0.105&108\\
\hline
\end{tabular}
}
\label{tableDR}
\end{table}

  Considering the four vowel pair samples separately,
  Hypothesis~\ref{Hypo1} on the population mean $\mathop{DR}$ holds at $p \leq
  0.05$ by fast rises only (Welch two sample (one-sided) t-test with
  $df=209$, $t = -1.8$, $p=0.04$).
%
The estimated mean and standard deviation of $\mathop{DR}$ using all
\textipa{[i]} glides are $0.320$ and $0.098$ ($n = 427$),
respectively. For all \textipa{[\textscripta]} glides, the respective
values are $0.337$ and $0.098$ ($n = 429$).  Hence,
Hypothesis~\ref{Hypo1} is verified using the full data set (Welch two
sample test with $df = 853$, $t=-2.5$, $p < 0.01$).
 
The first part of Hypothesis~\ref{Hypo2} is verified at $p = 0.05$ by
fast rise \textipa{[i]} (one-sided t-test, $df = 107$, $t = -3.1$,
$p=0.001$) and also by using all \textipa{[i]} glides as the sample
(one-sided t-test with $df = 426$, $t=-2.7$, $p < 0.01$).  The
population mean of $\mathop{DR}$ over all \textipa{[\textscripta]} is not
significantly different from the nominal value ($df=428$, $|t|<0.9$,
$p<0.4$) but the probability of a Type II Error is still large. Thus,
the latter part of Hypothesis~\ref{Hypo2} is not supported
statistically conclusively by the current dataset.

\begin{table}[t,h]
\label{tableRatioMixed}
\caption{\it The linear mixed-effects model results for Duration Ratio
  ($\mathop{DR}$). The fixed factors are vowel, direction and speed;
  the subjects are treated as random factors.}
\vspace{2mm}
\centerline{
\begin{tabular}{|c|c|c|c|c|}
\hline
Factor & Estimate & Std. Error & t-value & p-value\\
\hline
Intercept &  0.338 & 0.009 & 39.0 & 0\\
Vowel (\textipa{[i]}) & -0.017 & 0.007 & -2.5 & 0.01\\
Direction (rise) & -0.007 & 0.007& -0.98 & 0.3\\
Speed (slow) & 0.004  & 0.007 & 0.56 & 0.6\\
\hline
\end{tabular}
}
\end{table}

A mixed effects model supports the previous observations.  Glide
$j=1,\ldots,n_i$ from subject $i$ produces the $\mathop{DR}$ value denoted by
$y_{i,j}$, given by the mixed effects model
\begin{equation*}
y_{i,j} = \beta_0 + d_{i,j} \beta_1 + s_{i,j} \beta_2 + v_{i,j} \beta_3
	+  b_i + e_{i,j}
\end{equation*}
where the fixed factors (with numerical levels $0$ or $1$) are the
glide direction $d_{i,j}$, the glide speed $s_{i,j}$, and the vowel
type $v_{i,j}$. The test subject related random factors $b_i \sim
N(\mu_i,\sigma_i)$ are independent from the experimental errors
$e_{i,j} \sim N(0,\sigma)$.  As can be seen in the Table 2, the vowel
type has a significant effect (ANOVA, $df=845$, $|t|>2.5$, $p=0.011$)
while the effect of the direction and speed are not significant
($df=845$, $|t|<1$).  In a more complicated model taking into account
all the interactions of the factors, the log-likelihoods of the models
were lower, and the interactions were never significant.

\section{Discussion}

As is usual in model validation experiments, the measurement data is
much more ambiguous than the results of numerical simulations.  The
reconciliation of simulated and measured data is difficult because of
the following two main challenges:
\begin{enumerate}
\item \label{ValidationProblem1} The computational model could give an
  unrealistically strong indication of modal locking because of
  unmodelled physics (such as subglottal acoustics, supraglottal
  turbulence, and energy losses).
\item \label{ValidationProblem2} The humans have active control
  mechanisms (without counterparts in the computational model) that
  are expected to reduce the observable effects of modal locking.
\end{enumerate} 
Subglottal acoustics has been completely excluded from the
computational model.  The main reason for this is the lack of
experimental data that would be required to reproduce the extremely
difficult absorbing boundary condition that the progressively
subdividing system of bronchi and the alveoles constitute.  Moreover,
in the four-mass glottis model of Tokuda et al. \cite{TZKH:BMRTRVTR},
subglottal acoustics play a minor role compared to the supraglottal
acoustics. Hence, the coupling of the subglottal acoustics and the
vocal fold oscillations does not seem to be the primary source of the
discrepancy between simulations and experimental results.

The supraglottal aerodynamics might have influence on the vowel
glides.  State-of-the-art Computational Fluid Dynamics (CFD) models
indicate significant vorticity above the glottis
\cite{S-H-R:PCFDSF3DMVHVF}. The vorticity-induced aerodynamic force to
the vocal folds is a genuinely fluid mechanical feature that is beyond
the scope of the acoustic theory of speech as well as simple Bernoulli
flow models in VT. Evaluating the contribution of this force would
also require a more refined CFD model than is usually available.

In low frequency glide productions, the larynx moves vertically
\cite{H}. However, the simulated vowel glides were produced using a
fixed area function, and no movement in the larynx area is taken into
account at all.  The increased pitch is a consequence of the
elongation and higher stress in vocal folds due to rotation of the
thyroid cartilage in the anterior direction.  Towards the end of a
glide production, larynx moves as a consequence of both the
contracting thorax and the control actions required for variable
pitch. These two mechanisms either oppose or reinforce each other
depending on the direction of the glide.  All this applies equally to
VT geometries of \textipa{[\textscripta]} and \textipa{[i]}.

Considering the latter challenge~\ref{ValidationProblem2} above, the
direct modeling of the auditory, motor, and muscular mechanisms behind
the task performance seems unfeasible at the moment.  Fortunately, the
experiments can be arranged in a manner that downplays or even
interferes with the active control mechanisms. For this reason,
trained singers are -- somewhat paradoxically -- excluded as test
subjects even though producing vowel glides resembles very much a
singing exercise. (Recall that non-singing males were excluded for the
opposite reason.) The drawback of having non-singers was that some of
the subjects could not perform the glide task satisfactorily which may
have led to selection bias. However, these subjects are expected to be
less familiar in their vocal pitch control which would lead to less
accurate compensation (and hence, better detection in experiments) of
the modal locking disturbance.

Control mechanisms based on auditory feedback have been shown to
rapidly and accurately compensate deviations from target glides in
glide production \cite{K-J:SARSCAFWS}.  This could lead to asymmetric
$f_0$ trajectories in terms of the glide direction because of
different delays in control and observation. Recalling the simulated
$f_0$ trajectory in Fig.~\ref{fig:sweeps1}b, such a modal locking
event during a rising glide is detectable only right after a large
jump has already occurred while a modal-locking event during a fall
can, in principle, be detected when the $f_0$ has just ceased to
decline.  Considering the control delay alone, suppose that the
compensation strategy against modal locking were only based on the
auditory observation with an identical delay (of say 80 ms) for both
rising and falling glides.  Then the fall would be less perturbed by
the modal locking event because then the control action has more time
to react. For obvious reasons, however, the auditive observation of
modal locking events in falling glides could be more delayed compared
to rising glides for $f_0$ trajectory profiles as in
Fig.~\ref{fig:sweeps1}b.  All in all, there are convincing reasons why
rising and falling glides in experiments are not expected to be
symmetric, and such asymmetry could explain the fact that the fast
rising \textipa{[i]} stands out when verifying Hypothesis~\ref{Hypo2}
above.  It must be noted that the fast rising \textipa{[i]} does not
similarly stand out using the mixed effects model.

Not much can be concluded from the observed numerical values of
$\mathop{DR}$ on the two vowel classes. First, the distribution of
$F_1$ for \textipa{[i]} glides was not controlled while it is known by
the numerical experiment that a relatively small variation of $50$ Hz
in $\tilde F$ affects the mean values of $\mathop{DR}$.  Second,
$\mathop{DR}$ is a nonlinear functional, and one has to be careful in
interpretations by continuity; e.g., $\mathop{DR}$ may have very large
or very small values for large jumps that cross both $f_A$ and $f_B$.
Third, the jumps never occur instantaneously in the data but rather
show an accelerated pitch movement giving $\mathop{DR}$ values closer
to $1/3$ than in the simulations.  Fourth, the dynamics of the pitch
following exercise is not understood in such detail that its effect on
$\mathop{DR}$ could be evaluated at all. Such dynamics may be a source
of systematic error in the estimated $\mathop{DR}$ values, resulting
in the lack of verification of the latter part of
Hypothesis~\ref{Hypo2} concerning \textipa{[\textscripta]} glides.

The computational model predicts modal locking to subharmonics of the
vowel formants as well. These are characterized by jumps of one
octave, and, if they took place, they would have gone unnoticed in the
data processing. Indeed, the pitch detection algorithm admitted only
results inside one octave $[200 \, \mathrm{Hz}, 400 \, \mathrm{Hz}]$
which was the nominal gliding range.  Otherwise, large jumps were not
observed for \textipa{[\textscripta]} in the current dataset but such
jumps have been observed in other experiments
\cite{ZMHWH:OAIVVFTIPNSFCCS}.

\section{Conclusions}

Results from numerical simulations and experiments have been reported
on vowel glides.  The simulations show a robust and large effect of
modal locking of $f_0$ to $F_1$ whenever the frequencies coincide.
The experimental data shows a reduced effect of the same kind,
supporting the computational model. Hence, experimental results give
positive evidence for the existence of significant filter-source
feedback in human glide productions even in an experimental setting
where the frequency jumps were implicitly asked to be
avoided. Moreover, the results suggest that compensatory mechanisms
are employed to avoid filter-source feedback induced $f_0$ jumps near
the formant frequency $F_1$.

Uncontrolled $f_0$ jumps occur frequently in the phonation of boys
during the puberty \cite{BBZ:IVAPCNDV}.  Changes in the vocal folds
and the growing larynx make it difficult for them to produce natural
intonation contours or to sing musical melodies. These observations
may be partly due to modal locking since the relevant compensatory
mechanisms are expected to require re-tuning after a radical change in
the geometric dimensions of both the VT and the vocal folds.  If the
poorly compensated filter-source interaction were a significant cause,
then these $f_0$ jumps would have to be more frequent in VT
configurations where $F_1$ is low; i.e., in high vowels such as
\textipa{[i]} and \textipa{[u]}.

Finally, we remark that the active compensation mechanisms to
counteract the filter-source feedback could perhaps be observed
directly. It is expected that especially trained singers would 
adjust their VT in order to keep the formant and the vocal pitch apart
from each other.

\subsection*{Acknowledgements}
  
The authors were supported by the Finnish graduate school in
engineering mechanics, Finnish Academy grant Lastu 135005, the Finnish
Academy projects 128204 and 125940, European Union grant Simple4All
(grant no. 287678), Aalto Starting Grant, and {\AA}bo Akademi
Institute of Mathematics. The authors would like to thank O.~Engwall
for the area function used in the simulations.

\end{document}